\documentclass[aip,apl,amsmath,amssymb,reprint,superscriptaddress]{revtex4-1}
\usepackage{graphicx}
\usepackage{amsmath}
\usepackage{dcolumn}
\usepackage{bm}
\usepackage{bbold}

\bibstyle{apsrev4-1}

\begin{document}
    
\title{Ultrasensitive multi-mode ESR probed ferromagnetic two-level system of $Mn^{4+}$ impurity ion in the insulated $MnO_6$ complex of $SrLaAlO_4$ at $20~mK$}

\author{M. A. Hosain}
\email{akhter361@yahoo.co.uk}
\affiliation{ARC Centre of Excellence for Engineered Quantum Systems, School of Physics, University of Western Australia, 35 Stirling Highway, Crawley WA 6009, Australia.}



\begin{abstract}
   Ultrasnsitive multi-mode electron spin resonance spectroscopy in the $SrLaAlO_4$ dielectric resonator at $20~mK$ reveals ferromagnetic states of $Mn^{4+}$ impurity ion. The formation of ferromagnetic states in the $MnO_6$ complex implies to oxygen deficiency of this multi-valence $Mn^{4+}$ ion. Experiment results supports that an intricate electronic hybridization in $MnO_6$ structural instability is related to Pseudo Jahn-Teller effect. Measured dipolar hyperfine structure parameter of nucleus is $P_{\scriptscriptstyle\parallel} =-3.7\times 10^{-4}~cm^{-1}$. Mean inverse third power of the electron distance is $\langle r_q^{-3}\rangle=3.325~a.u.$ assuming nuclear electric quadruple moment $Q=+0.33(1)~barn$. In such a state, giant g-factor is observed due to magneto (ferromagnetic) impedance taking in to account a two-level system on the adiabatic-potential-energy-surface. The spins exhibited parity is opposite in the interaction of highest-occupied-molecular-orbital and lowest-unoccupied-molecular-orbital coupling. 
\end{abstract}

\maketitle

\subsection{Introduction:}

Ultrasensitive electron spin resonance (ESR) with high precision in dielectric crystal multi-mode resonators assimilating as hybrid-quantum system is a road-map for the development of quantum technologies$\cite{KurizkiHybSy,RoadMapHybSy,ContHybSy,MagnVibrSpinCrossBO}$. This experimental study is on impurity paramagnetic ion's unpaired electron spin states using electron spin resonance (ESR) spectroscopy in a suitable dielectric crystal resonator exciting microwave whispering gallery (WG) modes$\cite{Farr,Karim,Buluta}$. WG multi-mode ESR spectrum works as a direct probe providing information of electronic and magnetic states of paramagnetic impurity ions$\cite{SLA}$. In this process, $Mn^{4+}$ ion has been detected in the dielectric single crystal $SrLaAlO_4~(SLA)$, and analysed taking into account an intricate electronic hybridization due to its extra charge in $MnO_6$ complex at $20$ millikelvin $(mK)$. This $d-p$ metal-ligand orbital hybridization is mediated in $MnO_6$ structural instability, and plays a vital role in the mechanism of spontaneous polarization and/or magnetization forming two-level system on the adiabatic potential energy surface (APES)$\cite{JT_Review}$. Naturally, paramagnetic ion's three phenomena ferro-electricity, magnetization, and spin-crossover are observed as coexisted$\cite{MagnPJTEd0-d10}$. Ohkoshi et al.$\cite{MagnLightInduSpinCross}$ demonstrated unpaired electrons as a spin-crossover magnet in the mechanism of light induced phase transition. Such a transition process can be used to monitor magnetization saturation $(M_s)$, Curie temperature $(T_c)$, coercive (magnetic) field $(H_c)$ and/or the magnetic pole$\cite{MagnLightInduSpinCross}$. Optical spectroscopy and X-ray diffraction (XRD) results are available providing localization information of $Mn^{4+}$ ion in this type of crystal$\cite{SpectroscopyMn4,ElecMagCharSLA}$.\ 
 
  Many studies have been devoted towards better understanding of the main mechanisms governing electron delocalization and electron intervalence absorptions of transition metals in metal-ligand complexes$\cite{HamiltMMag,LocaDelocaPKS}$. These transition metal based crystals like $LaMnO_3$ exhibits interesting magnetic behaviours, such as, mono-metallic complexes in the crystal exhibiting single ion magnetic (SIM) behaviour$\cite{LocaDeloca,HamiltMMag}$, and two or more metal sites of varying oxidation numbers ( known as mixed valence (MV) sites$\cite{LocaDeloca,DefChemPerov}$ ) exhibits single molecular magnet (SMM) behaviour$\cite{HamiltMMag,DefChemPerov}$. This is essentially important due to the fact that these metal-ligand complex structures possess several potential applications in quantum technology$\cite{HamiltMMag,LocaDelocaPKS}$. However these characteristics due to their intrinsic magnetic properties, SMMs are distinctly detected only at liquid helium temperatures$\cite{HamiltMMag,LowTempGeomaFerro}$. Among the MV metal complexes, dual-exchange (DE)$\cite{LocaDeloca,HamiltMMag}$ is generated due to the presence of an itinerant electron in two different valence sites in the crystal. As instance, electron exchanges between two neighbouring sites $Mn^{3+}$ and $Mn^{4+}$ in the $Mn-O-Mn$ chain of $LaMnO_3$ crystal$\cite{Spinels}$. The MV metal ion complexes exhibit coupling of electron movement with the structural distortion, and subsequently affects the degree of localization of the extra electron$\cite{HamiltMMag,MnOxoGupta,MnValencePbESR}$. The oxidation variation of MV sites of the metal-ligand complex produces distortions as Jahn-Teller effect (JTE) within trigonal plane of lower symmetry, and can be confirmed from ESR spectrum$\cite{LocaDelocaPKS,MnOxoGupta,Spinels}$. \
    
 Intriguingly, we examine an insulated octahedral mono-metallic complex $MnO_6$ in $SrLaAlO_4$ where $Al^{3+}$ ion is substituted by $Mn^{4+}$ ion$\cite{DefChemPerov,SpectroscopyMn4}$. In this case the manganese ion shows multi-valence behaviour instead of MV behaviour. Interesting magnetic behaviours of this type of transition metal complex are observed in a linear combination of atomic orbitals. Neither the DE mechanism which is a type of magnetic exchange (whether materials are ferromagnetic or antiferromagnetic) that may arise between MV ions in the $Mn-O-Mn$ link, nor the super-exchange (SE) (or Kramers-Anderson super-exchange) which is a strong (usually) antiferromagnetic coupling between two next-to-nearest neighbour cations, is possible for an insulated $MnO_6$ individual unit in $SrLaAlO_4$. We will justify this multi-valance manganese ion in a metal-ligand charge transfer (oxidation variation) produced spontaneous magnetization with the measured spin-Hamiltonian parameters along with site symmetry$\cite{AbragamESR,CharlesESR,BleaneyPryce,CorrectSH}$.\ 
  
 The magnetic $Mn^{4+}$ ion in $MnO_6$ structure has a certain spin parity in the formed molecular orbitals$\cite{MagnPJTEd0-d10,HomoLumo}$. Here is our description for an appropriate realization of detected $Mn^{4+}$ ion's spin quantum state, which is dealt with ferromagnetism empirically rationalized to intricate electronic hybridization in $MnO_6$ complex referring to PJTE$\cite{MagnJT-ferro}$. Formation of two-level system is taken into account in the two minima on $APES$ in the mechanism of highest-occupied-molecular-orbital $(HOMO)$ and lowest-unoccupied-molecular-orbital $(LUMO)$ coupling.\cite{MagnJT-ferro,MagnPJTEd0-d10,HomoLumo,MagnMultiCross,MagnSymFerroFlexo}.  

\subsection{ESR spectroscopy experiment using WG modes:}

Field confinement of the WG mode in $SrLaAlO_4$ crystal allows loss mechanisms to be minimized to achieve a high Q-factor at $20~mK$\cite{Krupka,LeFloch,WG,Krupka1}, which is required for an ultrasensitive ESR spectroscopy. Using X-band to Ku-band frequency WG multi-mode ESR at this temperature, high precision is achieved in the measurements of the spin-Hamiltonian parameters$\cite{PilbrowESR,BleaneyIngram}$. Different process are devoted in measuring sensitivity of different type of resonator of wide range of frequency with varieties of probing system$\cite{AnninoQL,LongoQL,ColligianiQL,AndersMK,YapMK}$. Benmessai et al.$\cite{Karim}$ described a concentration level measurement process of $Fe^{3+}$ impurity ion exciting WG modes at millikelvin temperatures in sapphire. Anders et al.$\cite{AndersMK}$ described a single-chip electron spin resonance detector operating at $27~GHz$.\

For such a spectroscopy, a cylindrical light yellow $SrLaAlO_4$ single crystal of height $9.04~mm$ and diameter $17.18~mm$ was inserted centrally in an oxygen-free cylindrical copper cavity. The crystal loaded cavity was cooled in a dilution refrigerator (DR) to less than $20~mK$. Practically microwave-power and other terms are kept constant then the required minimum number of impurity ion follows the proportionality$\cite{CharlesESR}$ $N_{min} \propto \frac{1}{\omega Q_L}$ for detection of ESR transition spectrum. The required minimum spin number is estimated generally as:
 \begin{eqnarray}
\label{eq:Nmin}
 N_{min}=\Big(\frac{3 K_B V_sT_s}{g_e^2\beta^2\mu_\circ S(S+1)}\Big)\Big(\frac{\Delta \omega}{\omega}\Big)\Big(\frac{1}{\eta Q_L}\Big)\Big(\frac{P_n}{P}\Big)^\frac{1}{2}~~~~~~~   
\end{eqnarray} 
 Where $V_s$ is the mode volume, $T_s$ is the sample temperature, $S$ is the electron effective spin, $g_e$ is the electron g-factor, $\beta$ is the Bohr electron magneton, $\mu_\circ$ is the magnetic permeability of free space, $\omega$ is the resonance frequency, $\eta$ is the filling factor, $P_n$ is the noise power at the detector, $P$ is the microwave input power, and $\Delta \omega$ is the width of aggregated spin frequency at resonance which is depended on the shape-function $f(\omega)$ normalized as $\int_0^\infty f(\omega)\partial\omega=1$ for a wide range of Larmor precession $(\omega_L)$ of magnetic dipoles$\cite{AbragamESR}$. Significant output (transmission) occurs only at resonance in a very narrow frequency width $\Delta \omega$ in the region $\omega\approx\omega_L$ at ESR. In this experiment, for $Mn^{4+}$ ion $\Delta \omega\simeq 40~kHz$ (red band in the Fig.~\ref{MnDensityPlot}). The variation of Q-factor was small due to a little dielectric variation among the selected modes, and observed loaded Q-factor $Q_L$ was always more than 50,000 at $20~mK$. Assuming, $\Delta \omega$ in the order of the line-width of all the selected WG modes, the minimum number of detectible ions setting $\frac{P_n}{P}\approx1$ (see~Eq.$\ref{eq:Nmin}$) may be as low as $0.1~ppb$ level of concentration.\
 
 Fifteen WG modes with high-azimuthal-mode-number with a frequency range of $7~GHz$ to $18~GHz$, and thus electromagnetic energy filling factors of order unity were monitored. The static magnetic field between $-0.2~T$ to $1~T$ was varied through the use of computer control in a step of sweep $4\times 10 ^{-4}~T$. Each WG mode was scanned for a period of five seconds at each step of magnetic field sweep. This slow sweep of magnetic field was applied under control of an in-house MATLAB program to avoid heating above $20~mK$, with the microwave input power of $-60~dBm$.\

To avoid the addition of thermal noise from room temperature, a $10~dB$ microwave attenuator was used at $4~K$ stage and another one at $1~K$ stage of the DR. Also, a $20~dB$ attenuator was added at $20~mK$ stage of the DR. These cold stage attenuation plus the use of a low noise temperature cryogenic amplifier after the resonator ensures good enough signal to noise ratio $(SNR)$. From this multi-mode ESR characteristics with hyperfine structure plotting as a map ($Fig.\ref{Mn-g}$), we were able to identify the types of paramagnetic impurities present in the crystal.
             
\subsection{Results and Discussion}
\begin{figure}[t!]
\centering
\includegraphics[width=3.3in]{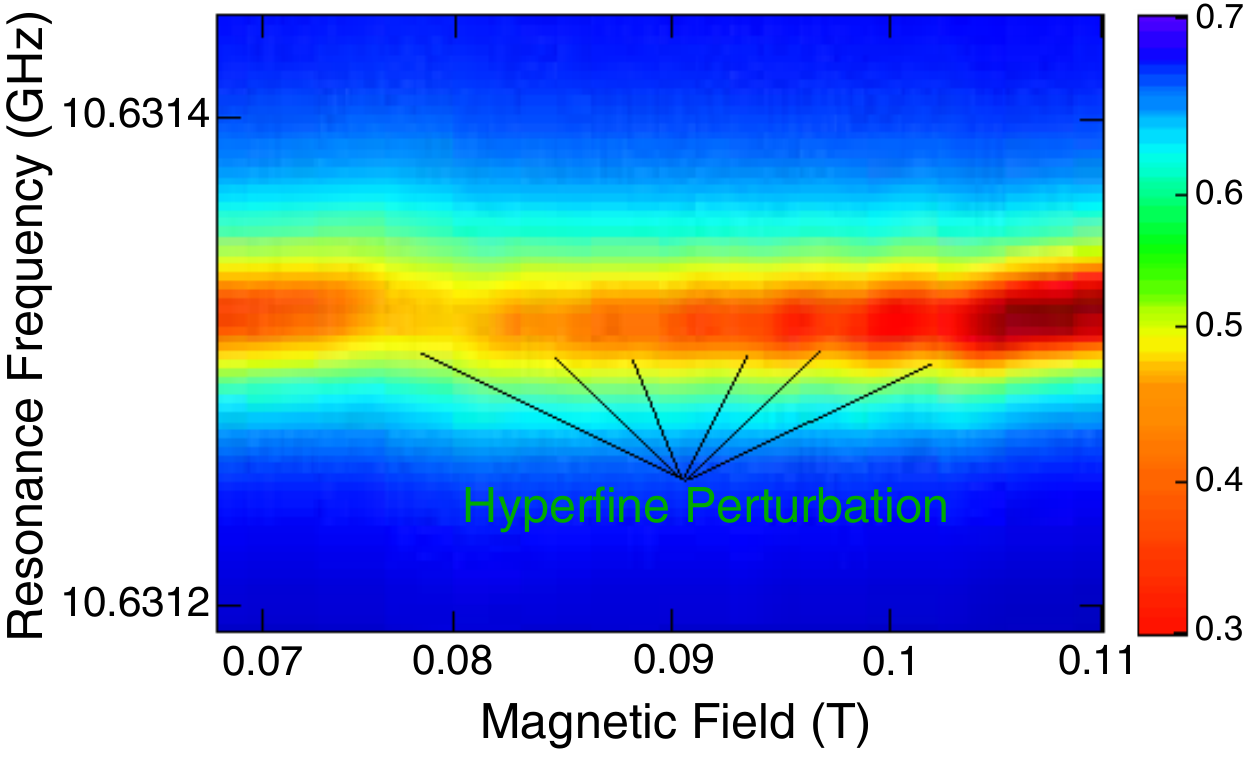}
\caption{\label{MnDensityPlot} Transmission spectrum colour density plot of $Mn^{4+}$ ion nuclear spin perturbation with $WGH_{5,1,1}$ mode of resonance frequency $10.6313~GHz$.}
\end{figure}

\begin{figure}[t!]
\includegraphics[width=3.5in, height=2.2in]{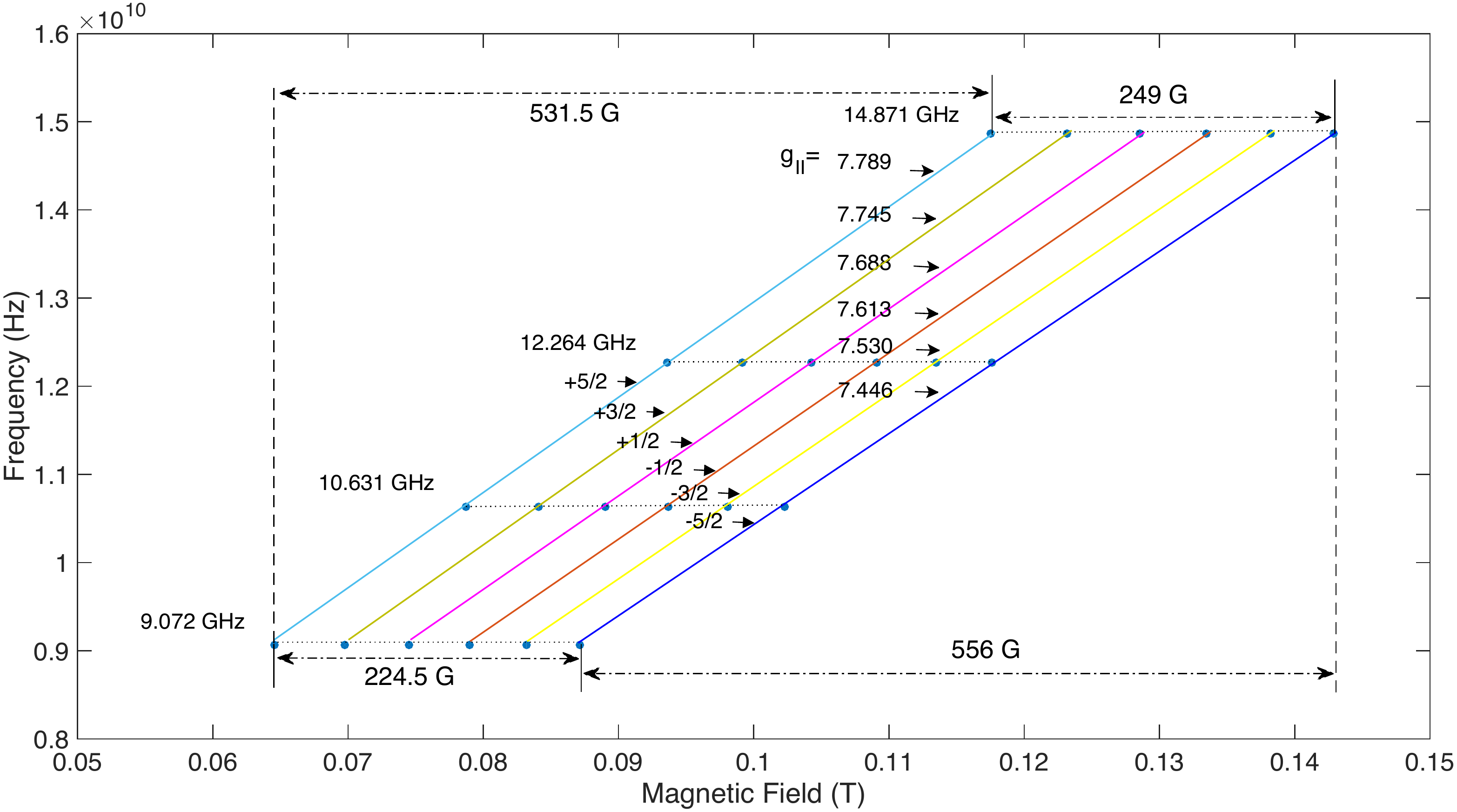}
\caption{\label{Mn-g} g-factor map of $Mn^{4+}$ ion ESR spectroscopy showing hyperfine structure broadening. Six lines of nuclear magnetic quantum numbers $+\frac{5}{2},+\frac{3}{2},+\frac{1}{2},-\frac{1}{2},-\frac{3}{2}$ and $-\frac{5}{2}$ shows different g-factors.}
\end{figure}

\begin{figure}[t!]
\includegraphics[width=3.1in]{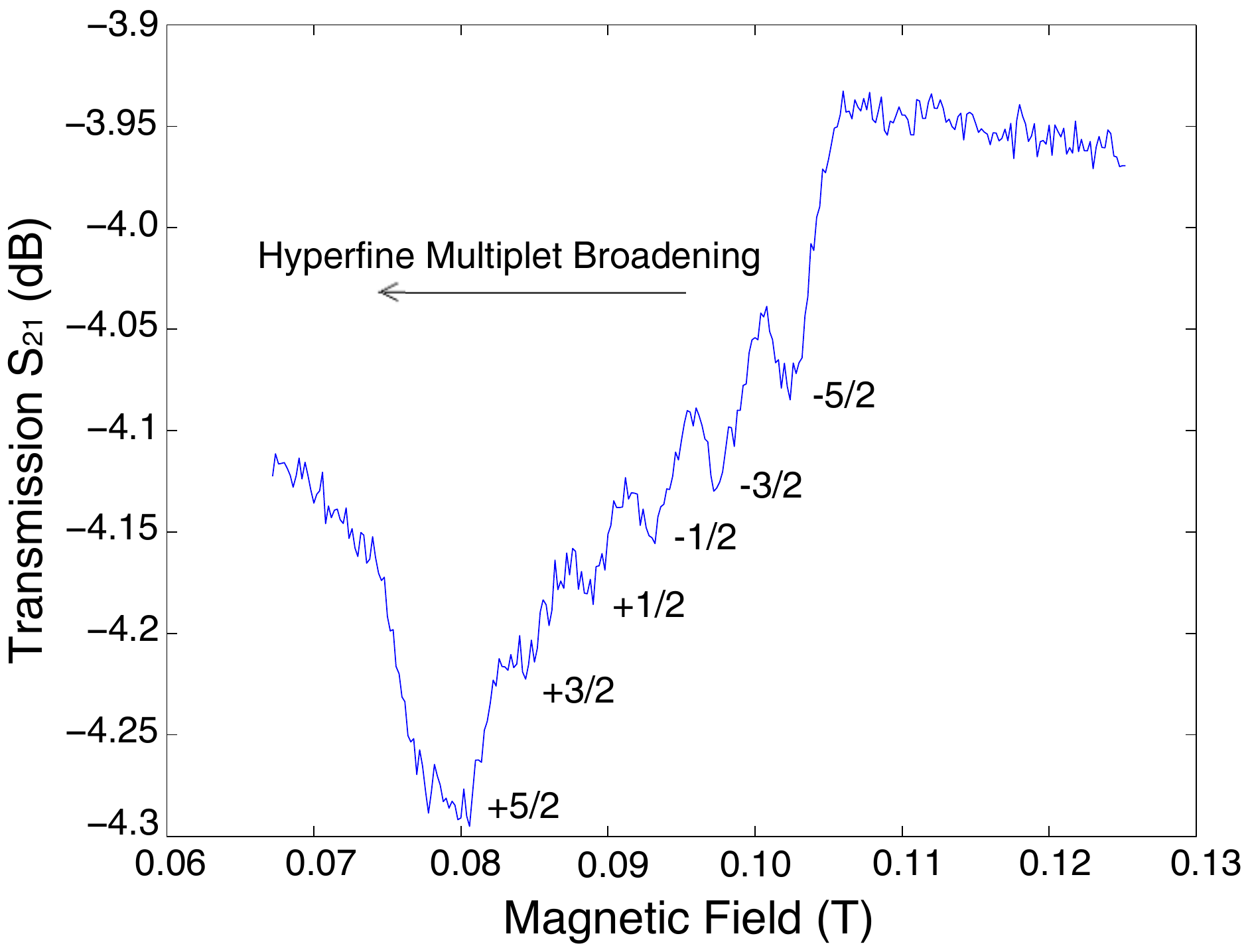}
\caption{\label{MnScratch} ESR spectrum of $WGH_{5,1,1}$ mode of resonance frequency $10.6313~GHz$. Spectrum of hyperfine line shows average spacing $44.16~G$ ($-178.5\times 10^{-4}~cm^{-1}$) of $Mn^{4+}$ ion at $20~mK$.}.
\end{figure}

Using the prescribed technique of experiment, the monitored ESR spectrum is mapped as in Fig-2. The isoelectronic $Cr^{3+}$ ion of $Mn^{4+}$ ion$\cite{CrMninMgO}$ has a nuclear spin $I=\frac{7}{2}$, but manganese has nuclear spin $I=\frac{5}{2}$ which is responsible for hyperfine structure of 6-lines ($Fig.~\ref{MnDensityPlot},\ref{Mn-g}$ and $\ref{MnScratch}$). The observed ESR spectrum assures the presence of $Mn^{4+}$ ion in the $SrLaAlO_4$ crystal lattice. Optical spectroscopy study of Zhydachevskii et al.$\cite{SpectroscopyMn4}$ shows that manganese ion is present exclusively in doped $SrLaAlO_4$ crystal tetragonal lattice in the form of $Mn^{4+}$ ion occupying six fold coordinated $Al^{3+}$ sites. Hence the presence of $Mn^{4+}$ ion in lower valence $Al^{3+}$ site is enhanced$\cite{ElecMagCharSLA}$. Also, the higher valence state of $Mn^{4+}$ ion resulting in stronger Coulomb interaction between $Mn^{4+}$ ion and $O^{2-}$ ion$\cite{SpectroscopyMn4,ElecMagCharSLA}$. The fact that in the ${^4}T_{2g}$ triplet, $Mn^{4+}$ ion has energy level overlapping depending on local charge density$\cite{SpectroscopyMn4,ElecMagCharSLA}$. This overlapping in the perovskite crystal SLA structure display variety of magnetic properties as a linear combinations of the atomic Hartree-Fock orbitals$\cite{HamiltMMag,LocaDelocaPKS,MnSpinOrdering,OrderDisorderXRD}$ in the molecular orbitals (MO) of $MnO_6$ complex$\cite{LocaDeloca,LocaDelocaPKS}$.

  Some anisotropy of $g$-factors and hyperfine line space broadening is observed, opposite to the direction of the increase of applied DC magnetic field ($Fig.~\ref{Mn-g}$ and $\ref{MnScratch}$). The geometrical anisotropy terms of the single $MnO_6$ structure is an important case where local order, as established by local interactions, cannot be freely propagated throughout space. The system can lift degeneracy resulting charge or spin ordering of manganese ion$\cite{CObeforeSO,SpinChOrdHF}$. Crystal distortion relates to Jahn-Teller distortion$\cite{SLACu}$, and metal-ligand charge transfer with orbital ordering plays an essential role in stabilizing ferromagnetic states$\cite{MagnPJTEd0-d10,MagnMultiCross,SpinChOrdHF}$.
 
   The measured parallel $g$-factors decreases in the order of nuclear magnetic quantum number $+\frac{5}{2},~+\frac{3}{2},~+\frac{1}{2},~-\frac{1}{2},~-\frac{3}{2},~-\frac{5}{2}$ with the increase of magnetic field ($Fig.~\ref{Mn-g}$ and $\ref{MnScratch}$). Measured parallel $g$-factors are $g_{\scriptscriptstyle\parallel Mn}= 7.789,~7.745,~7.688,~7.613,~7.530$ and $7.446$ according to the order $+\frac{5}{2}$ to $-\frac{5}{2}$. Similarly, hyperfine line spacings are $A_{\scriptscriptstyle\parallel Mn}=-209.8\times 10^{-4}~cm^{-1},~-196.4\times 10^{-4}~cm^{-1},~-182.7\times 10^{-4}~cm^{-1},~-169.1\times 10^{-4}~cm^{-1},~-159.7\times 10^{-4}~cm^{-1},~-153.7\times 10^{-4}~cm^{-1}$ according to the order of nuclear magnetic quantum number $+\frac{5}{2}$ to $-\frac{5}{2}$ at $10.6313~GHz~(WGH_{5,1,1})$ (Fig.~\ref{Mn-g} and \ref{MnScratch}).\
 
 The crystal field created large gap between $t_{2g}$ and $e_{2g}$ stabilizing $4+$ oxidation stage of manganese ion$\cite{DFT-Activators}$. It is a worthy remark that, in principle, any decrease in $MnO_6$ symmetry results in at least partial lifting of the orbital degeneracy, no matter how small the displacements are. Also, it has been observed by ESR in $PbTiO_3$ that valence state of doped manganese in $Ti^{3+}$ site changes from $Mn^{2+}$ to $Mn^{4+}$ with increase of its concentration$\cite{MnValencePbESR}$. This is usually accompanied by a distortion of crystal structure, typically through an interaction with the lattice$\cite{OOandFrust}$. Likewise any orbital degeneracy lifting in the crystallographic sites due to structural distortion is bound to entail differences in the total electron charges leading to a non-integer oxidation state. Plausibly, elongated $MnO_6$ octahedron due to tetragonal distortion in $SrLaAlO_4$ may implies on the modulation of charge density and variation of oxidation in the covalency state of the $Mn^{4+}$ ion extra charges in the substituted $Al^{3+}$ ion sites$\cite{MnOxoGupta,DFT-Activators}$. The tetragonal elongation (along c-axis) is $0.236$~\r{A} in the $Al^{3+}$ site oxygen octahedron of two bonds $Al-O2$ of length $R_\parallel = 2.121$~\r{A} and four coplanar bonds $Al-O1$ of length $R_\perp = 1.885$~\r{A} between aluminium and oxygen$\cite{Co-inSla,DefectCu^2}$. As an instance, it may be mentioned that about $100~cm^{-1}$ energy change in average can be caused by $0.01$~\r{A} off-center displacement of the impurity ion (due to structural distortion)$\cite{CuHole}$. This crystal may have a little rhombic distortion at $20~mK$ temperature and is not identified as a ferroelectric crystal.\  
 
\begin{figure}[h!]
\centering
\includegraphics[width=3.3in]{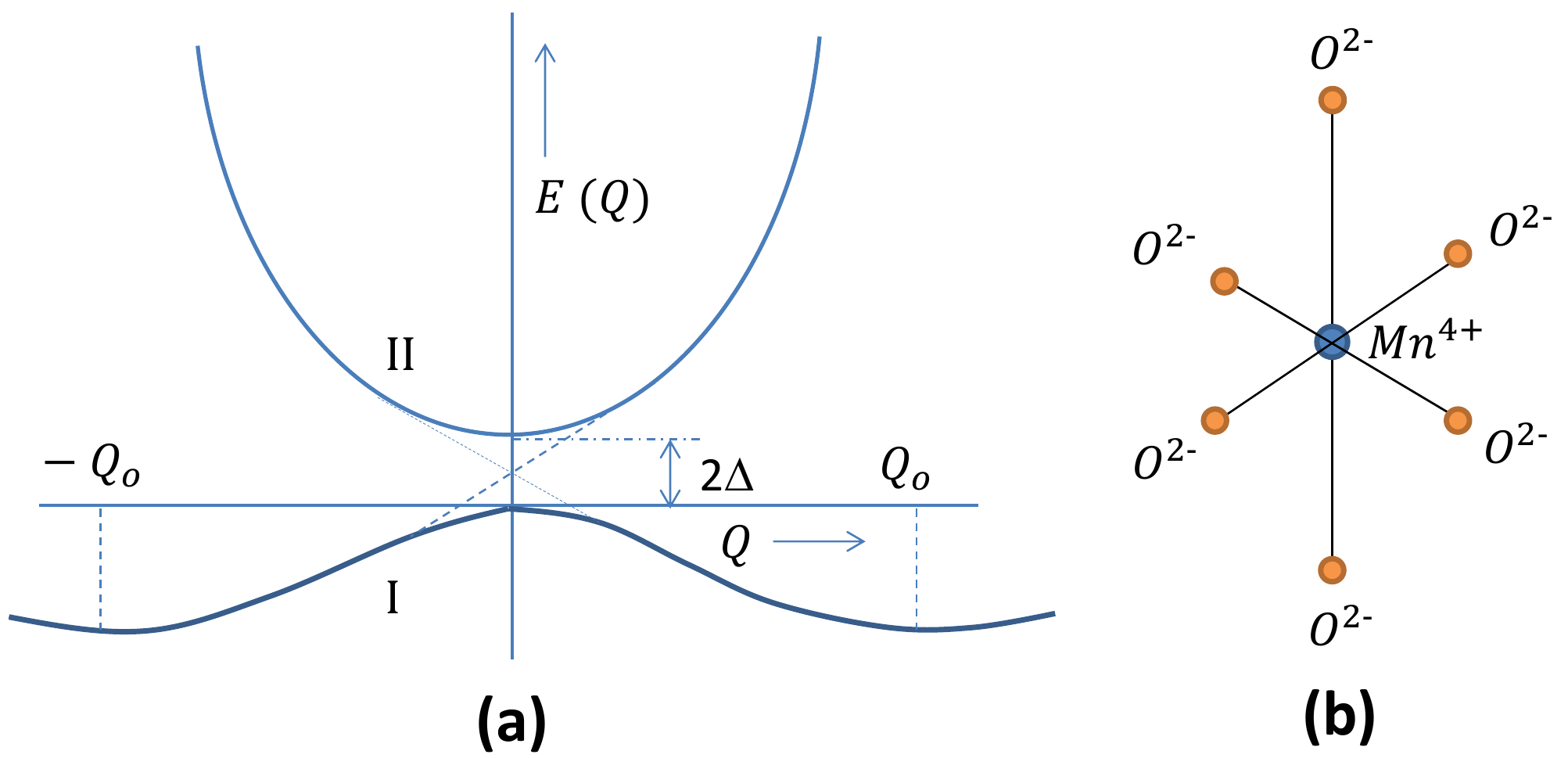}
\caption{\label{MnTwoLevel}- (a) Two-level system with two minima on APES at $+Q_\circ$ and $-Q_\circ$ due to PJTE at $\frac{F^2}{K_\circ} > \Delta$ state ($\Delta$ is the energy gap) of distortion in normal mode $Q=Q_\alpha (Q_x,Q_y,Q_z)$. At JTE stage the profiles are joining along dotted crossing lines: one from left to up right another from right to up left in a single distortion mode. (b) Elongated $MnO_6$ octahedral four-fold axes shape.}
\end{figure}

 Observed giant g-factor indicates high magnetic moment of electron in the $Mn^{4+}$ site at $20~mK$. ESR spectrum reveals this magnetization state directly as a spin-Hamiltonian parameter rationalizing to intricate electronic hybridization. In this hybridization, paramagnetic ion's three phenomena ferro-electricity, magnetization, and spin-crossover are observed as coexisted$\cite{MagnPJTEd0-d10}$. Empirically,  giant g-factor due to ferromagnetic two-level system which is formed in two potential minima on the APES between metal $Mn$ and ligand $O$ due to HOMO and LUMO coupling (Fig.\ref{MnTwoLevel}). The octahedral central manganese ion shifting with respect to oxygen $\bold{Q}_\alpha=(Q_x,~Q_y,~Q_z)$ in normal coordinates creates $MnO_6$ structural instability under a condition that the curvature of resultant spring constant $K=(\frac{\partial^2 E}{\partial Q_\alpha^2})_\circ$ (deviating from cubic symmetry) negative$\cite{MagnPJTEd0-d10,MagnJT-Green}$. $E=\langle \psi_\circ|H|\psi_\circ\rangle$ the energy at high-symmetry (cubic), the ground state wave function is $\psi_\circ$ and H is the metal-ligand interaction Hamiltonian.\
 
We consider that $K = K_\circ + K_v$. Where$\cite{MagnPJTEd0-d10}$;   
\begin{eqnarray}
\label{eq:K}
  K_\circ = \bigg\langle \psi_\circ \bigg |\bigg (\frac{\partial^2 H}{\partial Q_\alpha^2} \bigg )_\circ \bigg | \psi_\circ \bigg \rangle ~~~~~~~   
\end{eqnarray}   
 
 is the ground state diagonal matrix element. It describes the fixed (rigid) nucleus high symmetry electron density distribution $|\psi_\circ|^2$ reflecting stiffness of the lattice as a long-range (whole crystal) feature. Whereas, the term $K_v$ is always negative due to the Born-Oppenheimer ground state wave function and does not include long-range inter-cell interaction. This off-diagonal matrix elements are described in terms of second order perturbation as:$\cite{MagnJT-ferro,MagnJT-Green}$ 
 \begin{eqnarray}
\label{eq:Kv}
  K_v = -2\sum_n \frac{|\langle \psi_\circ | (\frac{\partial H}{\partial Q_\alpha} )_\circ | \psi_n \rangle |^2}{E_n - E_\circ} ~~~~~~~   
\end{eqnarray}     
 
 Instability arises in the structure under the condition $K_\circ + K_v <0$ in strong enough PJTE lower-symmetry due to manganese ions additional covalency with oxygen. We can consider  that $K_\circ$ and $K_v$ cancel one another approximately$\cite{MagnJT-Green}$. This allows us to focus on electronic interaction Hamiltonian for valence electrons only. In case of $MnO_6$ elongated (along z-axis) octahedron forming molecular orbitals as a linear combination of single electron Hartree-Fock wave function, the structure has attained the PJTE state of vibronic coupling at $20~mK$. The $HOMO$ is $t_{1u}$ energy level of oxygen $p_\pi$ function and $LUMO$ is $t_{2g}$ energy level of $Mn^{4+}$ ion $d_\pi$ function in hybridization$\cite{MagnJT-Green}$. Referring the wave functions $\psi_\circ$ and $\psi_n$ to HOMO and LUMO respectively, the vibronic coupling constant of PJTE for $Mn^{4+}$ ion additional covalency can be given as:$\cite{MagnJT-Green}{^,}\cite{MagnPJTEd0-d10,MagnJT-ferro}$
    \begin{eqnarray}
\label{eq:CouplingK}
  F = \bigg\langle 2p_z(O) \bigg| \bigg(\frac{\partial H}{\partial Q_x} \bigg)_\circ \bigg | 3d_{yz}(Mn) \bigg \rangle  ~~~~~~~   
\end{eqnarray}

This perturbation forms a two-level system in the profile of APES (Fig.\ref{MnTwoLevel}).\

\begin{figure}[h!]
\includegraphics[width=3.1in]{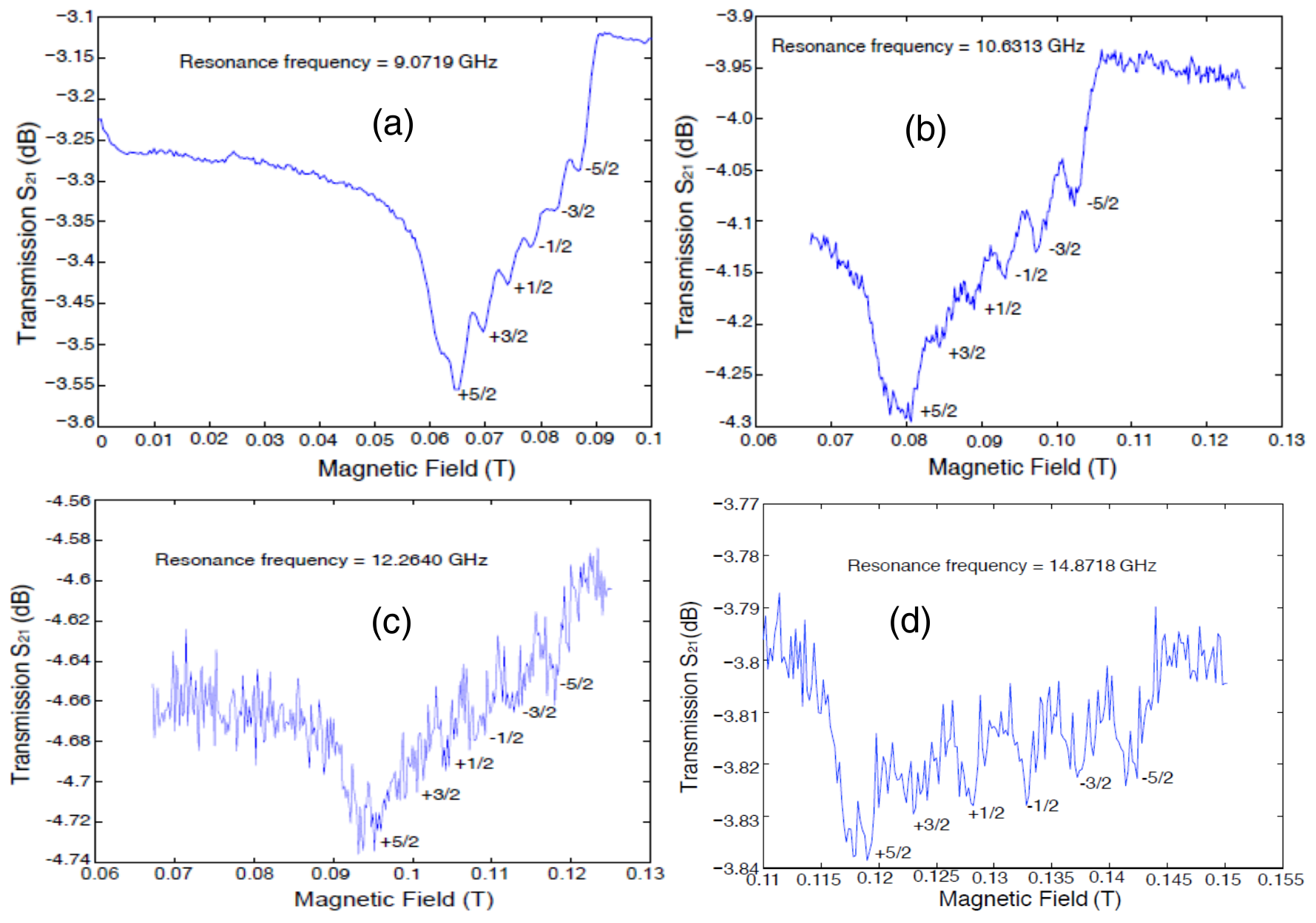}
\caption{\label{MnScratch2} Noisy hyperfine perturbation of $Mn^{4+}$ ion ESR spectrum due to magnetic impedance of mode frequency with the increase of frequency $9.0719~GHz$ to $14.8718~GHz$ arises in cascade due to coupling with ferromagnetic two-level system of higher local magnetization respectively. Also, slope of the electron transmission among consecutive nuclear spin perturbation is lowering with the increase of frequency. Figure-(a) shows clear hyperfine structure and ESR transmission spectrum, and stiff slope of electron transmission spectrum among consecutive nuclear spin hyperfine coupling.}
\end{figure}

The local character of the negative $K_v$ contribution to the curvature indicates that the instability producing PJTE is essentially of local origin, and the long range (whole crystal) interaction of $K_\circ$ is important in realization of instability condition $|K_v|>K_\circ$. This means that the PJTE of manganese ion center can be taken into account as two-level system$\cite{MagnJT-ferro}$ approximately reducing the denominator of the equation-$\ref{eq:Kv}$. Evidence of this instability reveals by the increased energy (or frequency) of spin transition in ESR due to higher magnetization of the ion $Mn^{4+}$ creating giant g-factor as observed in the experiment. In such a two-level state, higher frequency WG mode ESR transmissions should be noisy, and observed same results as shown in the figure-\ref{MnScratch2}a,b,c,d. The $d^3$ electron configuration of $Mn^{4+}$ ion effective spin state $S=\frac{3}{2}$ in the $t_{2g}$ orbital triplet is realised in the high crystal field of $SrLaAlO_4$. In contrast, according to the observed ESR spectrum of giant g-factor and low fine structure term, neither high-spin (HS) state $S=\frac{5}{2}$ nor low-spin (LS) state $S=\frac{1}{2}$ of $d^5$ electron of $Mn^{2+}$ ion is to be considerable in this substantially elongated $MnO_6$ octahedral structure at $20~mK$ temperature. Therefore, considering the typical molecular-orbital in energy scheme of $d^3$ electron spin configuration, the HOMO is $(t_{1u}\downarrow)^3(t_{1u}\uparrow)^3(t_{2g}\uparrow)^3$ with the ground state term ${^4A_{1g}}$ and the LUMO is $(t_{1u}\downarrow)^2(t_{1u}\uparrow)^3(t_{2g}\uparrow)^3(t_{2g}\downarrow)^1$ with the lowest ungerade term $^4T_{1u}$ of odd parity$\cite{MagnPJTEd0-d10,MagnJT-Green}$. This ground and excited states of opposite parity mediates the two-level vibronic coupling in the PJTE. An important feature is that it takes place as a magnetic dipolar effect in the $MnO_6$ complex.\

 The measured high parallel g-factor $g_{\scriptscriptstyle\parallel Mn}$ of $Mn^{4+}$ ion ESR in $SrLaAlO_4$ may compare with Suchocki et al.$\cite{GGG:Mn}$ observed enhanced zeeman effect with an effective g-factor in the range about 6 to 8 for $Mn^{4+}$ ion in gadolinium gallium garnet (GGG) at $25~mK$. It includes the degree of localization of the extra electron of $Mn^{4+}$ ion, and indicates that spin transition has magnetic impedance due to inherent 'frustration' coupling.\ 
 
  The excess positive charge of $Mn^{4+}$ increases its Coulomb force on the $O^{2-}$ surrounded site, and cause of orbital energy reduction. Meddey et al.$\cite{MagnFerroOxiVacIndu}$ claimed evidence of oxygen-vacancy induced ferromagnetic order of $Mn^{4+}$ ion in the Mn doped $SrTiO_3$ single crystal. Oxygen vacancy or deficiency $(\delta=0.15)$ in the sample is formed as $SrTi_{1-x}Mn_xO_{3-\delta}$, and its ferromagnetic phase shows a clear hysteresis loop at low temperature$\cite{MagnFerroOxiVacIndu,MagnFerroOxiVacElectron}$. Hence, the excess positive charge of $Mn^{4+}$ in the $MnO_6$ elongated octahedral sites of $SrLaAlO_4$ substituting $Al^{3+}$ having oxygen deficiency creates order-disorder state in ferromagnetic phase and interacts with external applied magnetic field of ESR spectroscopy. Typically, another example of oxygen vacancy mechanism for the reason of observed giant $g$-factor is as Gorni et al$\cite{Adduct}$ reported in their X-band parallel mode ESR experiment that the $g$-factor of manganese is 8.1 including $6.6~mT$ hyperfine structure line spacing at $5~K$, centered at $86~mT$. More clearly, the resonant frequency with applied external DC magnetic field $\bold{B}$ parallel to the crystal axis may be given by the Kittel formula\cite{Kittel1} $\omega ={\gamma }{\sqrt {B(B+\mu _{0}M)}}$. Where ${\displaystyle M}$ is the magnetization of the ferromagnet and ${\displaystyle \gamma }$  is the gyromagnetic ratio. As a result, determination of $g$-factor depends on the relative spin and orbital moments of a material which can be evaluated by use of the well-known relation\cite{Kittel1} $\frac{\mu_F}{\mu_S}=\frac{g-2}{2}$. Where, $\mu_F$ is the moment of spin in ferromagnetic state, and $\mu_S = \mu_B$ is the free spin moment.

Hence, we observed a giant g-factor as ESR is observed at higher resonance frequency with applied DC magnetic field B in addition with local magnetization M. In such a local environment with intricate electronic hybridization, measured nuclear hyperfine parameter is $P_{\scriptscriptstyle\parallel} \simeq -3.7\times 10^{-4}~cm^{-1}$ as an impact of nuclear second order perturbation (calculated using spin Hamiltonian). Using the value of manganese nuclear quadruple moment$\cite{NuclearQ}$ $Q=+0.33(1)~barn$, the measured mean inverse third power of the electron distance is $\langle r_q^{-3} \rangle = 3.325~a.u.$ at $20~mK$. According to the theory, this term is approximately $10-20\%$ higher in values for unfilled $d-$shell unpaired electrons.  Ionic radius of $Mn^{4+}$ ion in octahedral structure $r_{oh}=0.67$~\r{A} is in good agreement with these measurements$\cite{MnRadii}$.   
  
 Although, Prodi et al presented that with temperature, mixed-valence manganites with the $ABO_3$ perovskite structure display variation of properties with the relative concentration of $Mn^{3+}$ and $Mn^{4+}$ in the octahedral corner-sharing network$\cite{MnSpinOrdering}$. Instead of this mixed-valance sites, we may consider this case in the insulated single $MnO_6$ structure taking into account a non-integer oxidation state of $Mn^{4+}$ ion reducing from the $4+$ oxidation state. The spin still reasonably same as it was in case of $d^3$ electron configuration in orbital triplet with high energy gap between $e_g$ and $t_{2g}$. Although, spin-crossover mediates in orbital ordering resulting to structural change$\cite{MagnFerroEntOrboOrder}{^,}\cite{MagnDirHalfMeta}$. Without of spin-crossover, the orbital triplet splitting is viable for $d^3$ configuration taking into account only the excited (lifted) $|xz\rangle$ and $|yz\rangle$ states of $t_{2g}$ orbital triplet in the elongation along z-axis.\

It is observed that the hyperfine perturbation becomes noisy with the increase of resonance frequency ($Fig.\ref{MnScratch2}a,b,c,d$). Apparently, the magnetic impedance due to ferromagnetic order induces the electron spin transitions. Increase of magnetic reminiscence in the ferromagnetic hysteresis loop can deplete the distinction of spin interactions. Both the ferroelectric order and ferromagnetic order has hysteresis loop in the above mentioned metastable two-level system of order-disorder phase. Although, the experimental results reveals the impact of magnetic impedance but not able to identify the accurate scale of ferromagnetic two-level system of order-disorder.\ 
 
Regarding this two-level system, one electron Hamiltonian can be presented as $H=H_\circ + H_{P}$ with the part of PJTE perturbation $H_{P}$. In the second quantization formation due to $Mn^{4+}$ ion and $O^{2-}$ ion (HOMO and LUMO) electronic coupling, it is described with raising operator $a_i^\dagger (O_\mu)$ and lowering operator $a_j (O_\nu)$ as$\cite{MagnJT-Green}$- 
\begin{eqnarray}
\label{eq:H-pjte}
  H_{P} = \sum_{\alpha}\sum_{\mu,\nu}\sum_{i,j}\bigg \langle i,O_\mu \bigg |\frac{\partial H}{\partial Q_\alpha} \bigg |j, O_\nu \bigg \rangle Q_\alpha a_i^\dagger (O_\mu) a_j (O_\nu)  ~~~~~~  
\end{eqnarray}  

Where;  $\mu$, $\nu$ denotes different atoms and i, j denotes different orbitals of coupling. Also, $O_\mu$ and $O_\nu$ denotes octahedral symmetry of the orbitals.

\subsection{Conclusion:}
	
The elongated $MnO_6$ octahedral structure become unstable in lower symmetry raising metastable two-level system on the APES at $20~mK$. WG multi-mode ESR probes this instability directly revealing the $Mn^{4+}$ ion's valence electron states in situ. Structural anisotropy and oxygen deficiency has the vital role in formation of metastable two-level system in electronic hybridisation state which has been explained in terms of vibronic theory of HOMO and LUMO coupling. Hyper fine line width broadening measurement and covalent effect is important for microscopic state analysis at millikelvin temperatures revealing nuclear dipolar hyperfine parameter ($P_{\scriptscriptstyle\parallel}$) and mean inverse third power of the electron distance $\langle r_q^{-3} \rangle$. Measured value of $P_{\scriptscriptstyle\parallel}$ is negative as manganese nuclear electric quadruple moment is positive. Whereas, the value of $P_{\scriptscriptstyle\parallel}$ is positive for $Cu^{2+}$ ion in the same site as copper nuclear electric quadruple moment is negative$\cite{SLA}$.\

 Typically, in the formation of two-level system, bonding and anti-bonding mechanisms$\cite{LocaDeloca,LocaDelocaPKS}$ are involved in the linear combination of MO in the $MnO_6$ complex$\cite{HomoLumo,InvMnSH}$. WG mode ultrasensitive ESR spectroscopy probes these variety of intricate electronic local nature of magnetoelectric effects with high precision.

\begin{acknowledgements}

This work was funded by Australian Research Council (ARC), Grant no. CE110001013. Thanks to Dr. Warrick Farr for assistance with data acquisition and Mr. Steve Osborne for making cavity. 
\end{acknowledgements}

\subsection{References:}
\bibliography{biblio}
\end{document}